\documentclass[pdflatex,sn-mathphys]{sn-jnl}
\usepackage{txfonts}
\usepackage{here}
\usepackage{bm}
\usepackage{physics}
\graphicspath{{figures}}

\newcommand*{\filef}{pdf} 
\newcommand*{\fseco}{pdf} 

\newcommand*{\paren}[1]{\left( #1 \right)}
\newcommand*{\sbr}[1]{\left[ #1 \right]}
\newcommand*{\cbr}[1]{\left\{ #1 \right\}}
\newcommand*{\mrd}{\mathrm{d}}
\newcommand*{\sfbf}[1]{{\bfseries\sffamily #1}}
\newcommand*{\FT}{F_\mathrm{T}}
\newcommand*{\FN}{F_\mathrm{N}}
\newcommand*{\muM}{\mu_\mathrm{M}}
\newcommand*{\muS}{\mu_\mathrm{S}}
\newcommand*{\muK}{\mu_\mathrm{K}}
\newcommand*{\vc}{v_\mathrm{c}}
\newcommand*{\ve}{v_\mathrm{e}}
\newcommand*{\Pext}{P_\mathrm{ext}}
\newcommand*{\Vdri}{V}
\newcommand*{\sigmaf}{\sigma^\mathrm{(fric)}}
\newcommand*{\Stilde}{\tilde{S}}
\newcommand*{\Sc}{\tilde{S}_\mathrm{c}}
\newcommand*{\Azero}{A_0}
\newcommand*{\etat}{\eta_\mathrm{t}}
\jyear{2021}%
\theoremstyle{thmstyleone}%
\theoremstyle{thmstyletwo}%
\theoremstyle{thmstylethree}%
\begin{document}

\newcommand{\articletitle}{Control of Static Friction by Designing Grooves on Friction Surface}
\title[\articletitle]{\articletitle}

\author*[1]{\fnm{Wataru} \sur{Iwashita}}\email{w\_iwashita@fm.me.es.osaka-u.ac.jp}

\author[2]{\fnm{Hiroshi} \sur{Matsukawa}}

\author[1]{\fnm{Michio} \sur{Otsuki}}

\affil*[1]{\orgdiv{Department of Mechanical Science and Bioengineering}, \orgname{Osaka University}, \orgaddress{\street{1-3 Machikaneyama}, \city{Toyonaka}, \postcode{560-8531}, \state{Osaka}, \country{Japan}}}

\affil[2]{\orgdiv{Department of Physical Sciences}, \orgname{Aoyama Gakuin University}, \orgaddress{\street{5-10-1 Fuchinobe}, \city{Sagamihara}, \postcode{252-5258}, \state{Kanagawa}, \country{Japan}}}

\abstract{
This study numerically investigated the friction of viscoelastic objects with grooves.
A 3D viscoelastic block with grooves on a rigid substrate is slowly pushed from the lateral side under uniform pressure on the top surface. 
The local friction force at the interface between the block and the substrate obeys Amontons’ law. 
Numerical results obtained using the finite element method reveal that the static friction coefficient decreases with increasing groove width and depth. 
The propagation of the precursor slip is observed before bulk sliding.
Furthermore, bulk sliding occurs when the area of slow precursor slip reaches a critical value, which decreases with increasing groove size. 
A theoretical analysis based on a simplified model reveals that the static friction coefficient is related to the critical area of the precursor, which is determined by the instability of the precursor. 
A scaling law for the critical area is theoretically predicted, and it indicates that the decrease in the effective viscosity due to the formation of the grooves leads to a decrease in the static friction coefficient. 
The validity of the theoretical prediction is numerically confirmed. 
}


\maketitle

\clearpage
\subsection*{Graphical Abstract}
\begin{figure}[H]
    \centering
    \includegraphics[scale=0.7]{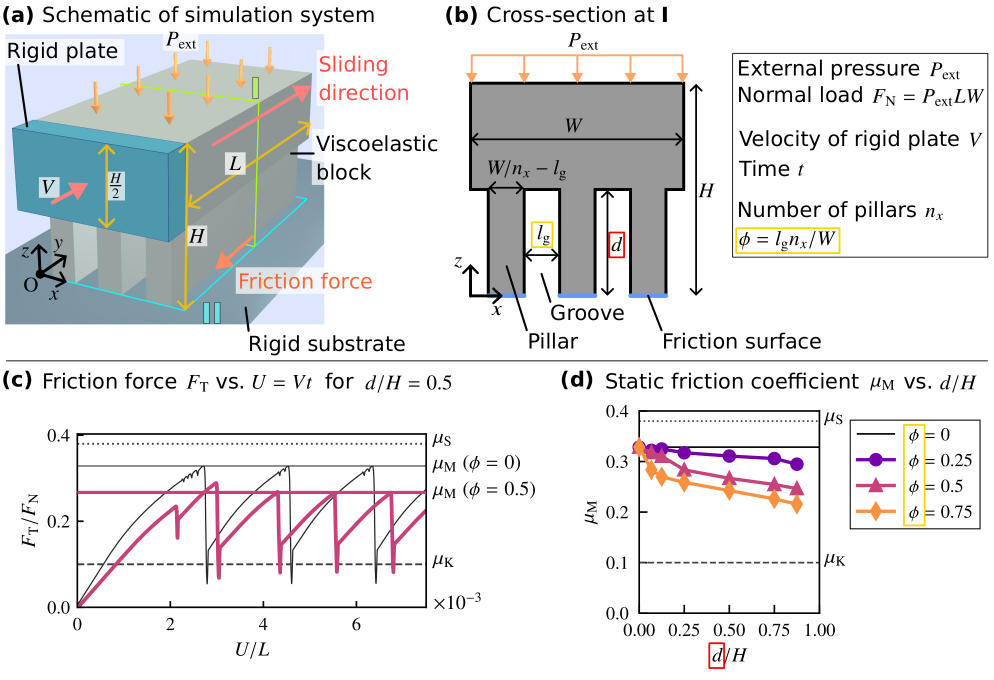}
\end{figure}

\section{Introduction} \label{sec:intro}

Friction forces occur in different situations, such as sliding parts of machines and contact surfaces between tires and the ground, and prevent the relative motion between two objects in contact. 
Friction forces are desirable in applications requiring low slippage, whereas they are undesirable in sliding parts of machines due to energy loss. 
Therefore, the control of friction forces is important in engineering~\cite{Bowden1950, Persson2000, Popov2017_text, Rabinowicz1995, Dowson1998, Bhushan2013, Baumberger2006}. 
One of the known methods to control friction forces is designing friction surfaces by forming grooves.
Generally, grooves on the surfaces of tires and sliding parts of machines are formed to reduce undesirable lubrication in wet conditions, which leads to a decrease in the friction coefficient and results in accidental slippage~\cite{Li2004, Li2005, Li2006, Yamaguchi2012}. 
However, the dependence of friction force on grooves in dry conditions has not been established clearly. 

Generally, for friction between solids in dry conditions, Amontons’ law is expected to hold~\cite{Bowden1950, Persson2000, Popov2017_text, Rabinowicz1995, Dowson1998, Bhushan2013, Baumberger2006}. 
According to Amontons’ law, the friction coefficient does not depend on the external pressure or size and shape of the object. 
However, the phenomenological explanation of Amontons’ law is based on the adhesion of microscopic asperities at the friction interface~\cite{Bowden1950, Persson2000, Popov2017_text, Rabinowicz1995, Dowson1998, Bhushan2013, Baumberger2006, Archard1957, Dieterich1996}, and implicitly assumes the uniformity of the stress field. 
For macroscopic objects associated with the non-uniform stress field, Amontons’ law may not hold. 
Therefore, the friction coefficient may depend on the shape of the macroscopic objects in dry conditions. 

In fact, recent studies on the friction of objects with flat friction surfaces have shown that Amontons’ law is not satisfied when the non-uniformity of the stress field is significant~\cite{Bouissou1998, Ben-David2011, Otsuki2013, Katano2014, IwashitaSciRep2023}.
In Refs.~\citenum{Otsuki2013, IwashitaSciRep2023}, numerical simulations and analysis of a simplified model have revealed the mechanism of breakdown of Amontons’ law in viscoelastic materials. 
The analysis clarified that the local precursor slip before bulk sliding due to the non-uniform stress field leads to the breakdown of Amontons’ law, and that the static friction coefficient exhibits characteristic load dependence. 
The relationship between the precursor slip and breakdown of Amontons’ law and the load dependence of the static friction coefficient have been verified in experiments on acrylic glass blocks~\cite{Katano2014}. 
Precursor slip relates to earthquake~\cite{Selvadurai2017, Yamaguchi2011, Obara2016, Kato2021, Petrillo2020, Xu2023} and fracture~\cite{Svetlizky2014, Bayart2016, Svetlizky2017, Berman2020, Gvirtzman2021, Kammer2015, Castellano2023}, and has been extensively studied in experiments~\cite{Katano2014, Selvadurai2017, Yamaguchi2011, Xu2023, Rubinstein2007, Ben-David2011, Svetlizky2014, Bayart2016, Svetlizky2017, Berman2020, Gvirtzman2021, Maegawa2010} and numerical simulations~\cite{IwashitaSciRep2023, Petrillo2020, Maegawa2010, Burridge1967, deSousa1992, Braun2009, Capozza2012, Ozaki2014, Scheibert2010, Amundsen2012, Tromborg2014, Radiguet2013, Kammer2015, Castellano2023, Albertini2021, Taloni2015, deGeus2019}. 
However, many of these studies have considered only flat friction surfaces, and the effects of grooves in the friction surface on frictional properties and precursor slip are yet to be discovered. 

Recently, several studies have been conducted to reveal the dependence of the friction coefficient on the macroscopic shape of the friction surface. 
In Refs.~\citenum{Capozza2015, Costagliola2016, Costagliola2017, Costagliola2018, Costagliola2022IJSS, Maegawa2017}, shapes of friction surfaces were represented by a spatial dependence of the local friction coefficient in 1D or 2D spring-block models. 
The frictional properties of the models vary with the spatial pattern of the local friction coefficient~\cite{Capozza2015, Costagliola2016, Costagliola2017, Costagliola2018, Costagliola2022IJSS, Maegawa2017}, and these results have been applied to experiments of macroscopic objects~\cite{Berardo2019, Balestra2022}. 
However, it is unclear to what extent the results of the spring-block models with a spatial pattern of local friction coefficient reflect the effect of the actual surface shape. 
Experiments with rubber and gel blocks have also revealed the dependence of the friction coefficient on the macroscopic shape of the friction surface~\cite{Maegawa2016, Gao2023}.
However, it is unclear whether the results for relatively soft objects such as rubber and gel can be applied to harder materials where Amontons’ law is locally satisfied. 

In this study, using the finite element method (FEM), we numerically investigate the friction of a 3D viscoelastic material with grooves in a dry condition, where the friction force locally obeys Amontons’ law. 
The dependence of the static friction coefficient on the groove shape is investigated. 
We find that the static friction coefficient is a decreasing function of the groove width and depth.
We also observe that local precursor slip occurs before bulk sliding of the viscoelastic material. 
The bulk sliding occurs when the area of the precursor slip reaches a critical value. 
The static friction coefficient is scaled by the normalized critical area of the precursor slip. 
The propagation of the precursor slip is analytically studied based on a simplified model. 
We derive the conditions for the onset of bulk sliding and the dependence of the static friction coefficient on the groove shape.
The results show that the static friction coefficient decreases due to the decrease in effective viscosity as the groove width and depth increase. 
Both pillars in the friction surface and main body supporting them play an important role. 
Our results aid in the improvement of sliding interface design by making grooves for both wet and dry conditions.

\section{Model and Methods} \label{sec:model_methods}

\begin{figure}[b]
    \centering
    \includegraphics[scale=0.7]{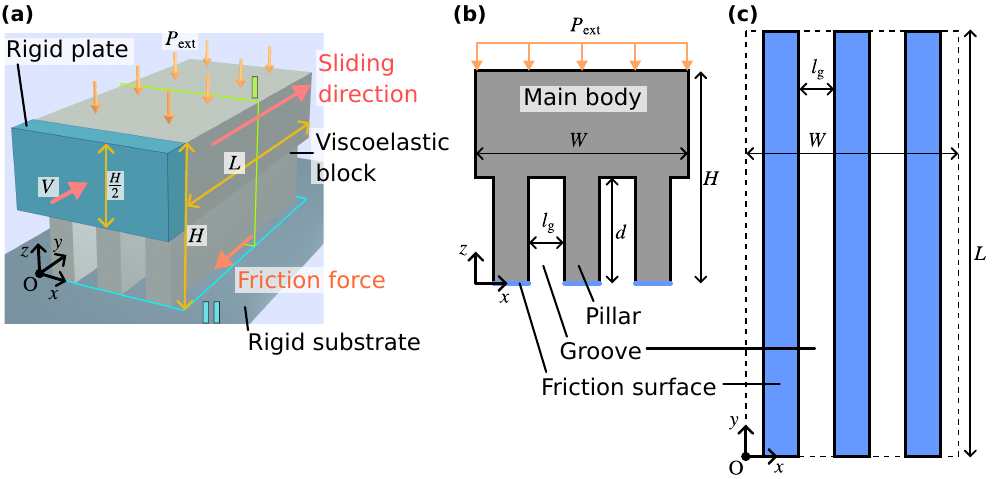}
    \caption{
        Schematic of the system.
        (a)~Grooved viscoelastic block moving on a rigid substrate.
        (b)~Cross-section perpendicular to the $y$ direction indicated as \sfbf{I} in (a). 
        (c)~The bottom of the block indicated as \sfbf{II} in (a). 
        The blue region represents the contact area between the rigid substrate and the block. 
    }
    \label{fig:schem_fem}
\end{figure}

We consider grooved viscoelastic blocks on a rigid substrate under a uniform external pressure $\Pext$ with width $W$, length $L$, and height $H$ along the $x$, $y$, and $z$ axes, respectively, as shown in Fig.~\ref{fig:schem_fem}. 
The rigid substrate is at $z=0$. 
A rigid plate with a width of $W$ and a height of $0.5 H$ pushes the side of the block at $y=0$ and $0.5 H \leq z \leq H$ with a slow constant velocity $\Vdri$ along the $y$ direction. 
This study considers a longitudinal groove parallel to the $y$ direction, as shown in Fig.~\ref{fig:schem_fem}. 
The number of pillars in the friction surface of the block is denoted by $n_x$. 
The pillars are equally spaced with width $l_\mathrm{g}$. 
The height and width of the pillar are denoted by $d$ and $W/n_x - l_\mathrm{g}$, respectively. 
The cross-section perpendicular to the $y$ direction is symmetrical, as shown in Fig.~\ref{fig:schem_fem}b. 
The pillar is in contact with the rigid substrate, as shown in Fig.~\ref{fig:schem_fem}c. 
The ratio $\phi$ of the area of the non-contact surface to the area of the bottom surface is given by $\phi = l_\mathrm{g} n_x / W$, and the contact area of the friction surface is given by $LW(1-\phi)$. 
Here, $\phi=0$ corresponds to a rectangular block without grooves. 

The equation of motion for the viscoelastic object is given by 
\begin{equation}
    \rho \ddot{u}_i = \sum_j \partial_j \sigma_{ij}
    \label{eq:eom_block}
\end{equation}
with density $\rho$, displacement vector $\bm{u}$, and stress tensor $\bm{\sigma}$; where $\sigma_{ij}$ is the $ij$ component of $\bm{\sigma}$, ${u}_i$ is the $i$ component of $\bm{u}$, and $\ddot{u}_i$ is its second-order time derivative. 
We adopt the Kelvin-Voigt model for $\bm{\sigma}$, where $\bm{\sigma}$ is given by $\bm{\sigma} = \bm{\sigma}^\mathrm{(E)} + \bm{\sigma}^\mathrm{(V)}$ with the elastic stress $\bm{\sigma}^\mathrm{(E)}$ obeying Hooke’s law and the viscous stress $\bm{\sigma}^\mathrm{(V)}$ proportional to strain rate, which reduces the elastic waves caused by the deformation of the block. 
We assume that the viscoelastic material of the block is isotropic.
The $ij$ component of the elastic stress tensor $\sigma^{ ( \mathrm{E} ) }_{ij}$ is given by 
\begin{equation}
    \sigma^\mathrm{(E)}_{ij} = \frac{E}{1+\nu} \epsilon_{ij} + \frac{\nu E}{(1+\nu)(1-2\nu)}  \sum_{k}\epsilon_{kk} \delta_{ij}
    \label{eq:sigma_E}
\end{equation}
with Young’s modulus $E$, Poisson’s ratio $\nu$, Kronecker’s delta $\delta_{ij}$, and strain tensor $\epsilon_{ij}$. The $ij$ component of the viscous stress tensor $\sigma^{ ( \mathrm{V} ) }_{ij}$ is given by 
\begin{equation}
    \sigma^\mathrm{(V)}_{ij} = \eta_1 \dot{\epsilon}_{ij} + \eta_2 \sum_{k}\dot{\epsilon}_{kk} \delta_{ij}
    \label{eq:sigma_V}
\end{equation}
with the two viscosity coefficients $\eta_1$ and $\eta_2$ and the strain rate tensor $\dot{\epsilon}_{ij}$~\cite{Landau1986}. 

The boundary conditions on the top surface of the block at $z=H$ are given by $\sigma_{zz} = -\Pext$ and $\sigma_{xz} = \sigma_{yz} = 0$. 
At surfaces except the top and bottom of the block, free boundary conditions ($\sum_j \sigma_{ij} n_j = 0$) are applied, where $n_j$ is the $j$ component of the normal vector $\bm{n}$ to the surface. 
The boundary conditions at the contact surface with the rigid plate at $y=0$ are given by $\sigma_{xy} = \sigma_{zy} = 0$ and $\dot{u}_y = \Vdri$, where $\dot{u}_y$ is the velocity along the $y$ direction. 
We set $\Vdri$ sufficiently small to push the block quasi-statically. 

The friction between the block bottom and substrate obeys Amontons’ law locally. 
Since the substrate is rigid, the $z$-direction displacement $u_z$ satisfies $u_z \ge 0$. 
At the bottom, the tangential stress vector $\bm{t}(x,y) = (\sigma_{xz}, \sigma_{yz})$ at the position $(x, y)$ is given by 
\begin{equation}
    \bm{t} = -\frac{\bm{v}}{v} \, \sigmaf \ ,
    \label{eq:vector_t}
\end{equation}
\begin{equation}
    \sigmaf(x, y) = \mu(v(x, y)) \, p(x, y) \ ,
    \label{eq:sigma_fric}
\end{equation}
where $\sigmaf$ is the frictional stress, $\bm{v}(x,y) = (\dot{u}_x, \dot{u}_y)$ is the slip velocity vector with velocities along the $i$ direction $\dot{u}_i$, and $v(x,y) = \lvert \bm{v} \rvert$ is the slip velocity~\cite{ccm2006}. 
The bottom pressure $p(x,y) = - \sigma_{zz}(x, y, z=0)$ is set to satisfy $u_z \geq 0$, where $p=0$ for $u_z > 0$. 
Here, $\mu(v)$ is the local friction coefficient depending on $v$. 
In the static region with $v(x,y)=0$, $\mu(v)$ is lower than $\muS$, and set to balance the local internal shear stress with the frictional stress. 
In the slip region with $v(x, y) > 0$, $\mu(v)$ is given by 
\begin{equation}
    \mu (v) = \left\{
    \begin{array}{ll}
        \muS - \paren{ \muS -\muK } v/\vc, ~& 0 < v < \vc \\
        \muK, ~& v \geq \vc
    \end{array}
    \right. \ ,
    \label{eq:mu_v_ver1}
\end{equation}
where $\muS$ and $\muK$ are the local static and dynamic friction coefficients, respectively. 
Here, $\vc$ is the characteristic velocity. 
The local Amontons’ law is expected to hold when a local region considered in the interface contains a sufficiently large number of real contact points, and has a negligibly small spatial variation in internal stress~\cite{Archard1957, Dieterich1996, Dieterich1994}. 
Note that the rate and state-dependent friction law~\cite{Baumberger2006} might be more appropriate to represent the local friction, but it coincides with the velocity-weakening friction law in Eq.~\eqref{eq:mu_v_ver1} for a sufficiently large slip length, which is satisfied in the poly methyl methacrylate (PMMA) experiments~\cite{Baumberger1999, Rubinstein2007, Katano2014, Bayart2016}. 
Hence, we have adopted Eq.~\eqref{eq:mu_v_ver1}. 
For blocks without grooves, the analysis using the velocity-weakening friction law~\cite{Otsuki2013} has been shown to reproduce the PMMA experimental results~\cite{Katano2014}. 

We numerically solve Eq.~\eqref{eq:eom_block} using FEM. 
The viscoelastic block is divided into cubes with length $\Delta x$, comprising six tetrahedrons. 
The displacements of its nodes are evolved based on Eq.~\eqref{eq:eom_block}, and the displacement and velocity within each element are approximated using linear interpolation. 
The local friction coefficient $\mu(v)$ is approximately given as 
\begin{equation}
    \mu (v) = \left\{
    \begin{array}{lll}
        \muS\,v/\ve, ~& 0 \leq v \leq \ve \\
        \muS - \paren{ \muS -\muK } v/\vc, ~& 
        \ve < v < \vc \\
        \muK, ~& v \geq \vc
    \end{array}
    \right.
    \label{eq:mu_v}
\end{equation}
with a sufficiently small velocity scale $\ve$. 
The state with $0 \leq v \leq \ve$ corresponds to the static region, and the state with $v > \ve$ corresponds to the slip region. 
We set $\ve/\Vdri = 2.5\times10^{-2}$ to satisfy $\ve \ll \Vdri, \vc$, and use $\Delta x/H=1/48$, $\Delta t v_\mathrm{s}/H \thickapprox 10^{-6}$, and $\Vdri/ v_\mathrm{s} =2.83\times 10^{-5}$ with $v_\mathrm{s} = \sqrt{E/\rho}$. 
Here, $v_\mathrm{s}$ represents the elastic wave velocity.
We set the driving speed $V$ to satisfy the condition $V << v_\mathrm{s}$, where the elastic waves are sufficiently dissipated. 
We have confirmed that the dependence of the results on the driving velocity $V$ is negligible under the condition of $V \ll \vc \ll v_\mathrm{s}$. 

In our simulation, we first apply a uniform pressure $\Pext$ to the top surface and relax the system to an equilibrium state. 
From the time $t=0$ after the relaxation, the rigid plate pushes the side of the block with a constant velocity $\Vdri$, and the calculation continues until a periodic stick-slip is observed. 

We set the length and width of the block to $L/H=4$ and $W/H=1$, respectively. 
Qualitatively similar results are obtained for $L/H=2$, as shown in Appendix~\ref{apsec:dependence_L}. 
We adopt $\nu=0.34$, $\eta_1 v_\mathrm{s}/( H E ) = 2.83$, $\eta_2/\eta_1=1$, $\muS=0.38$, $\muK=0.1$, and $\vc/v_\mathrm{s} = 4.81\times10^{-4}$ following previous simulations~\cite{Otsuki2013, IwashitaSciRep2023}. 
We select the number of pillars in the friction surface as $n_x=3$, and confirm that the dependence of the numerical results on $n_x$ is small, as shown in Appendix~\ref{apsec:dependence_nx}. 
In this study, we investigate the dependence on the external pressure $\Pext$, fraction of non-contact area $\phi$, and groove depth $d$.

\section{Results} \label{sec:results}

\subsection{Numerical Simulation} \label{subsec:fem_simulation}

\begin{figure}[b]
    \centering
    \includegraphics[scale=0.7]{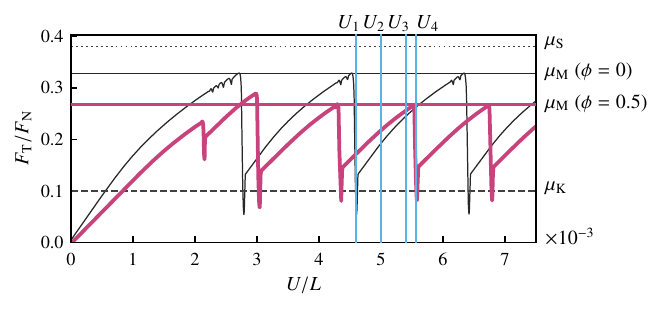}
    \caption{
        Ratio of friction force $\FT$ to applied normal force $\FN$ against displacement of the rigid plate $U$ for $\Pext/E=0.003$ and $d/H=0.5$. 
        The thin and thick solid lines represent the results for $\phi=0$ and $\phi=0.5$, respectively. 
        The thin and thick horizontal solid lines represent macroscopic static friction coefficient $\muM$ for $\phi=0$ and $\phi=0.5$, respectively. 
        The dotted and dashed lines represent $\muS$ and $\muK$, respectively. 
    }
    \label{fig:friction_force}
\end{figure}

Figure~\ref{fig:friction_force} shows the friction force $\FT$ against the displacement of the rigid plate $U = \Vdri t$ at time $t$ for $\Pext/E = 0.003$ and $d/H = 0.5$. 
Here, $\FT$ is given by the force on the rigid plate in the $y$ direction. 
In Fig.~\ref{fig:friction_force}, $\FT$ is normalized by the normal load $\FN = \Pext LW$ applied to the top surface of the block. 
The thin and thick solid lines represent the results for $\phi=0$ and $\phi=0.5$, respectively. 
For each $\phi$, $\FT/\FN$ increases approximately linearly with $U$, and rapidly decreases after reaching a peak value. 
When the rapid decrease occurs, the entire system slides, and the block returns to a static state after reaching a minimum value close to the local dynamic friction coefficient $\muK$. 
The increase and decrease in $\FT/\FN$ repeat periodically, which corresponds to stick-slip motion. 
We define the maximum value of $\FT/\FN$ in the periodic stick-slip region as the macroscopic static friction coefficient $\muM$, which is lower than the local static friction coefficient $\muS$. 
Figure~\ref{fig:friction_force} shows that $\muM$ for the block with grooves is lower than that for the flat block. 

\begin{figure}[b]
    \centering
    \includegraphics[scale=0.7]{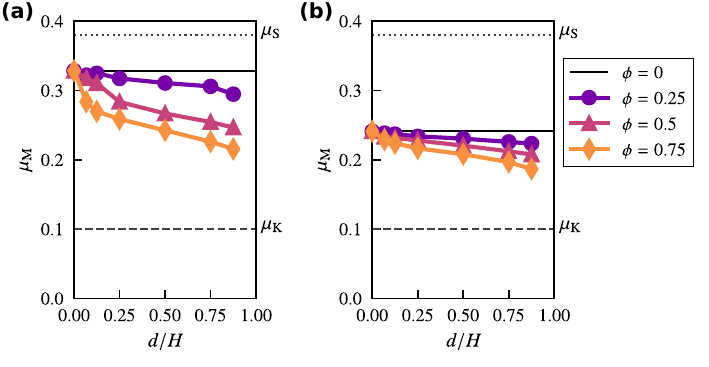}
    \caption{
        Macroscopic static friction coefficient $\muM$ against $d$ for different values of $\phi$ with (a)~$\Pext/E=0.003$ and (b)~$\Pext/E=0.006$. 
        The dotted and dashed lines represent $\muS$ and $\muK$, respectively. 
    }
    \label{fig:muM_d}
\end{figure}

In Fig.~\ref{fig:muM_d}, we plot the macroscopic static friction coefficient $\muM$ against the groove depth $d$ for different values of $\phi$ with $\Pext/E=0.003$ and $0.006$. 
Note that the results for $\phi=0$ are independent of $d$. 
For each $\Pext$, $\muM$ is a decreasing function of $d$. 
As $d$ approaches 0, $\muM$ converges to that for $\phi=0$. 
The macroscopic static friction coefficient $\muM$ is a decreasing function of $\phi$. 
These results indicate that the static friction force decreases as the size of the groove increases. 
Comparing Fig.~\ref{fig:muM_d}a and b, we find that $\muM$ is a decreasing function of $\Pext$, which is consistent with the results of previous studies on rectangular blocks without grooves~\cite{Otsuki2013, Katano2014, IwashitaSciRep2023}. 

\begin{figure}[b]
    \centering
    \includegraphics[scale=0.7]{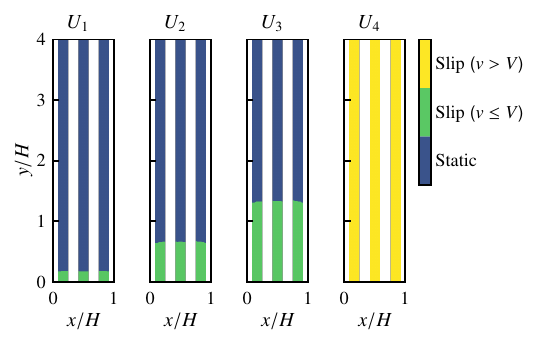}
    \caption{
        Spatial distributions of the slip velocity $v$ in the friction surface at $z=0$ for $U=U_1, U_2, U_3$, and $U_4$ shown in Fig.~\ref{fig:friction_force} for $\Pext/E = 0.003$, $\phi=0.5$, and $d/H=0.5$. 
        The blue area represents the static region. 
        The yellow-green and yellow areas represent the slip regions with $v \leq V$ and $v > V$, respectively. 
        The rigid plate pushes the block at $y=0$.
        The white area represents the groove region. 
    }
    \label{fig:cmap_precursor}
\end{figure}

In Fig.~\ref{fig:cmap_precursor}, we present the spatial distributions of the slip velocity $v$ in the friction surface at $z=0$ for the displacements $U=U_1, U_2, U_3$, and $U_4$ shown in Fig.~\ref{fig:friction_force} with $\Pext/E = 0.003$, $\phi=0.5$, and $d/H=0.5$. 
Here, we select $U_1/L=4.6\times10^{-3}$, $U_2/L=5\times10^{-3}$, $U_3/L=5.4\times10^{-3}$, and $U_4/L=5.57\times10^{-3}$ in the periodic stick–slip region. 
In Fig.~\ref{fig:cmap_precursor}, the blue area represents the static region,
and the yellow-green and yellow areas represent the sliding regions with $v\leq V$ and $v> V$, respectively. 
The quasi-static precursor slip with $v \leq V$ begins to propagate from the region near the rigid plate at $y=0$ for $U=U_1$, and the area of precursor slip expands quasi-statically as $U$ increases to $U_2$ and $U_3$. 
After $U_3$, the area of precursor slip develops rapidly, and the entire system begins to slide, leading to bulk sliding at $U_4$ (see Supplementary Videos). 
During bulk sliding, the slip velocity $v$ exceeds $V$. 
We confirm that the displacement due to these slips is approximately along the $y$ direction for all parameters. 

\begin{figure}[b]
    \centering
    \includegraphics[scale=0.7]{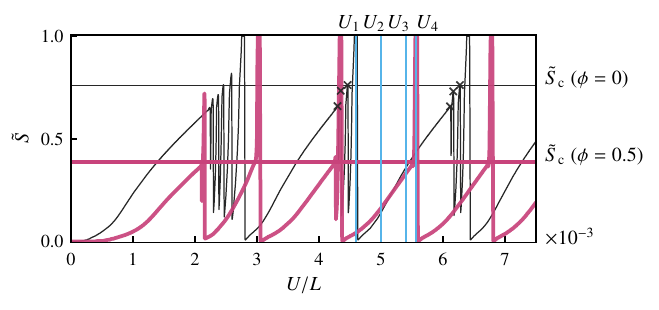}
    \caption{
        Normalized precursor slip area $\Stilde$ against $U/L$ for $\Pext/E=0.003$ and $d/H=0.5$.
        The thin and thick lines represent the results for $\phi=0$ and $\phi=0.5$, respectively. 
        The thin and thick horizontal lines represent the normalized critical area of precursor slip $\Sc$ for $\phi=0$ and $\phi=0.5$, respectively. 
        Crosses represent the peaks of the oscillation of $\Stilde$ in the region of the periodic stick-slip motion for $\phi = 0$. 
    }
    \label{fig:stilde_U}
\end{figure}

Figure~\ref{fig:stilde_U} shows the normalized slip area $\Stilde=S/[LW(1-\phi)]$ against $U/L$ for $\Pext/E = 0.003$ and $d/H=0.5$. 
Here, the precursor slip area $S$, defined by the sum of the yellow-green and yellow areas in Fig.~\ref{fig:cmap_precursor}, is normalized by the contact area in friction surface $LW(1-\phi)$. 
When $\Stilde=0$, the entire friction surface is static, while $\Stilde=1$ indicates bulk sliding where the entire friction surface is sliding. 
The thin and thick solid lines represent the results for $\phi=0$ and $\phi=0.5$, respectively. 
The normalized precursor slip area $\Stilde$ increases gradually with $U$ for small $\Stilde$, but the oscillation of $\Stilde$ appears as $\Stilde$ becomes large.
The slip associated with the oscillation in $\Stilde$ is called bounded rapid precursor (BRP). 
In BRP, the slip front propagates close to the elastic wave speed, but the slip quickly slows down and stops. 
According to our analysis in Sect.~\ref{subsec:theory_analysis}, the BRP is caused by oscillatory instability. 
The oscillation of $\Stilde$ in Fig.~\ref{fig:stilde_U} and small drops of $\FT/\FN$ in Fig.~\ref{fig:friction_force} before bulk sliding is caused by the sequence of the BRP~\cite{Otsuki2013}. 
Each BRP reduces the stress and $\Stilde$, but they both recover quickly due to a slight increase in the driving force. 
The BRP becomes significant depending on the values of the parameters. 
When $\Stilde$ reaches a threshold value $\Sc$, the propagation speed of $\Stilde$ suddenly increases, and $\Stilde$ reaches unity, which corresponds to the bulk sliding. 

We evaluate the critical slip $\Sc$ for bulk sliding in the periodic stick-slip motion as the maximum value of $\Stilde$ in the sequence of the BRP.
For example, we have plotted the peaks of $\Stilde$ due to BRP in the region of periodic stick-slip motion, $U/L > 3$, as crosses for $\phi = 0$ in Fig. \ref{fig:stilde_U}. 
The last peaks before bulk sliding represent $\Sc$, which are shown as horizontal lines in Fig. \ref{fig:stilde_U}. 
Figure~\ref{fig:stilde_U} shows that $\Sc$ decreases with increasing $\phi$.

\begin{figure}[b]
    \centering
    \includegraphics[scale=0.7]{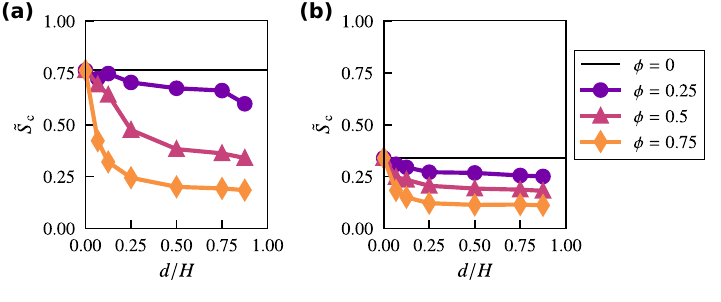}
    \caption{
        Normalized critical area of precursor slip $\Sc$ against $d$ for different values of $\phi$ with (a)~$\Pext/E=0.003$ and (b)~$\Pext/E=0.006$. 
    }
    \label{fig:sc_d}
\end{figure}

In Fig.~\ref{fig:sc_d}, we show the normalized critical area of the precursor slip $\Sc$ against the groove depth $d$ for different values of $\phi$ with $\Pext/E=0.003$ and $0.006$. 
We find that $\Sc$ is a decreasing function of $d$. 
As $d$ approaches 0, $\Sc$ approaches that for $\phi=0$. 
We also find that $\Sc$ decreases with increasing $\phi$. 
Comparing Fig.~\ref{fig:sc_d}a and b, we see that $\Sc$ is a decreasing function of $\Pext$, which is consistent with the results of previous studies on blocks without grooves~\cite{Otsuki2013, Katano2014, IwashitaSciRep2023}. 

\begin{figure}[t]
    \centering
    \includegraphics[scale=0.7]{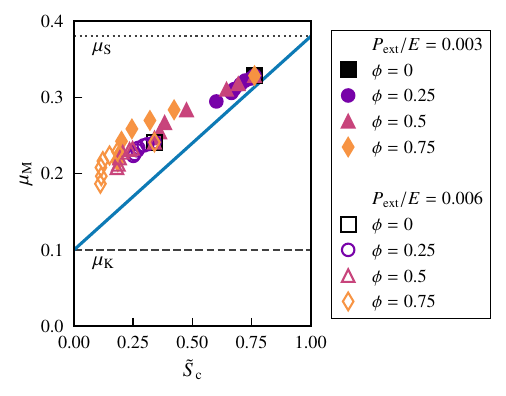}
    \caption{
        Macroscopic static friction coefficient $\muM$ against $\Sc$ for different values of $\phi$ and $d$. 
        The filled and open symbols represent the results for $\Pext/E=0.003$ and $\Pext/E=0.006$, respectively. 
        The solid line represents the analytical results given by Eq.~\eqref{eq:muM_Sc}. 
        The dotted and dashed lines represent $\muS$ and $\muK$, respectively. 
    }
    \label{fig:muM_Sc_tp_L}
\end{figure}

The dependence of the macroscopic static friction coefficient $\muM$ on $\phi$ and $d$ shown in Fig.~\ref{fig:muM_d} is similar to that of $\Sc$ shown in Fig.~\ref{fig:sc_d}. 
This similarity indicates a close relation between $\muM$ and $\Sc$. 
In fact, as shown in Fig.~\ref{fig:muM_Sc_tp_L}, $\muM$ is an almost linear function of $\Sc$ for different values of $\phi$ and $d$ with $\Pext/E=0.003$ and $0.006$. 
Figure~\ref{fig:muM_Sc_tp_L} also shows that $\muM$ lies between $\muS$ and $\muK$. 
This scaling of $\muM$ using $\Sc$ is consistent with the results of previous studies on blocks without grooves~\cite{Otsuki2013, Katano2014, IwashitaSciRep2023}. 

\begin{figure}[t]
    \centering
    \includegraphics[scale=0.7]{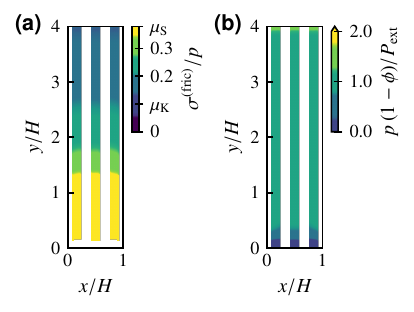}
    \caption{
        Spatial distribution of stress in the friction surface at $U=U_3$ for $\Pext/E = 0.003$, $\phi=0.5$, and $d/H=0.5$. 
        (a)~Spatial distribution of ratio of frictional stress $\sigmaf$ to bottom pressure $p$. 
        (b)~Spatial distribution of $p$. 
        The rigid plate pushes the block at $y=0$.
        The white area in (a) represents the region without contact. 
        The white area in (b) represents the groove area. 
    }
    \label{fig:cmap_tp}
\end{figure}

Figure~\ref{fig:cmap_tp}a shows the spatial distribution of the ratio of the frictional stress $\sigmaf$ to the bottom pressure $p$ in the friction surface at $U=U_3$ for $\Pext/E = 0.003$, $\phi=0.5$ and $d/H=0.5$. 
Here, $U=U_3$ indicates the state just before bulk sliding, as shown in Figs.~\ref{fig:friction_force} and \ref{fig:stilde_U}. 
Note that the local static friction in the static state with $v = 0$ takes any values for $0<\sigmaf/p<\muS$. 
As shown in Supplementary Videos and previous studies~\cite{Otsuki2013, IwashitaSciRep2023}, the ratio $\sigmaf/p$ returns to the value near $\muK$ in the entire area just after the bulk sliding. 
As the block is pushed, $\sigmaf/p$ reaches the local static friction coefficient $\muS$ near the region pushed by the rigid plate. 
The area with $\sigmaf/p \thickapprox \muS$ gradually increases as $U$ increases. 
The region with $\sigmaf/p \thickapprox \muS$ corresponds to the slip region at $U=U_3$ in Fig.~\ref{fig:cmap_precursor}, while $\sigmaf/p$ remains near $\muK$ in the static region. 

Figure~\ref{fig:cmap_tp}b shows the spatial distribution of the bottom pressure $p$ at $U=U_3$ for $\Pext/E = 0.003$, $\phi=0.5$ and $d/H=0.5$. 
Although a uniform pressure $\Pext$ is applied at the top surface, the spatial average of $p$ becomes $\Pext/(1-\phi)$, because the contact area in the friction surface, $LW(1-\phi)$, is smaller than the area of the top surface, $LW$, due to the grooves. 
We confirm that the bottom pressure is $p \approx \Pext/(1-\phi)$ in most areas except for the regions near $y=0$ and $L$. 
The spatial distribution of $p$ is almost independent of the time $t$, as shown in Supplementary Videos. 

\subsection{Theoretical Analysis} \label{subsec:theory_analysis}

\begin{figure}[b]
    \centering
    \includegraphics[scale=0.7]{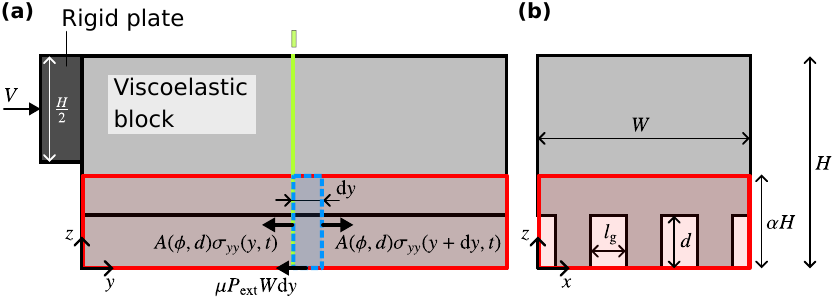}
    \caption{
        (a)~Schematic of the derivation of the simplified model for grooved viscoelastic block. 
        (b)~Cross-section perpendicular to the $y$ direction indicated as \sfbf{I} in (a).
        The red shaded areas represent the region from the bottom to the height $z = \alpha H$. 
        The dotted rectangle represents the element with infinitesimal width $\mrd y$. 
    }
    \label{fig:schematic_model}
\end{figure}

We theoretically analyze the effect of the longitudinal grooves shown in Sect.~\ref{subsec:fem_simulation} based on a simplified model~\cite{Otsuki2013, IwashitaSciRep2023}. 
The precursor slip is approximately uniform in the $x$ direction and propagates toward the $y$ direction, as shown in Fig.~\ref{fig:cmap_precursor}. 
Therefore, we neglect displacements in the $z$ and $x$ directions and consider only the $y$-dependent displacement along the $y$ direction. 
Since the bottom pressure $p$ at $z=0$ is approximately uniform, as shown in Fig.~\ref{fig:cmap_tp}b, we assume $p = \Pext/(1-\phi)$. 
Additionally, since the deformation is significant in the region near the bottom before bulk sliding in our 3D simulations, as shown in Appendix~\ref{apsec:acc_cross-section}, we focus on the slip and deformation in the region $0 \leq z/H \leq \alpha$ with a constant $\alpha$, as shown in the red shaded area in Fig.~\ref{fig:schematic_model}. 

We consider the equation of motion for a thin element at $y$ with small width $\mrd y$ indicated by the dotted rectangle in Fig.~\ref{fig:schematic_model}a. 
The mass of the element is given by $\rho A(\phi, d) \mrd y$, where $A(\phi, d)$ is the cross-sectional area of the red region in Fig.~\ref{fig:schematic_model}b excluding the groove. 
In Fig.~\ref{fig:schematic_model}a, the forces acting on the left and right surfaces of that element are given by $A(\phi, d) \sigma_{yy} (y, t)$ and $A(\phi, d) \sigma_{yy} (y + \mrd y, t)$, respectively. 
Here, the normal stress in the $y$ direction is denoted by $\sigma_{yy}$. 
The friction force acting on the bottom is given by $\mu \Pext W \mrd y$. 
The equation of motion for the displacement $q_y(y,t)$ of the thin element along the $y$ direction is given by 
\begin{equation}
    \rho A(\phi, d) \mrd y \, \ddot{q}_y (y, t) = A(\phi, d) \sbr{ \sigma_{yy}(y + \mrd y,t) - \sigma_{yy}(y,t) } - \mu(\dot{q}_y (y,t)) \Pext W \mrd y \ ,
    \label{eq:eom_ldl}
\end{equation}
where $\dot{q}_y(y, t)$ and $\ddot{q}_y(y, t)$ are the first and second-order time derivatives of $q_y(y,t)$, respectively. 
We assume a plane stress state, where the normal stress $\sigma_{yy}(y,t)$ is given by 
\begin{equation}
    \sigma_{yy}(y,t) = E_1 \frac{\partial q_y(y,t)}{\partial y} + \etat \frac{\partial \dot{q}_y (y,t)}{\partial y}
    \label{eq:sigma_yy}
\end{equation}
with the elastic modulus $E_1 = E/[(1+\nu)(1-\nu)]$ and viscous modulus $\etat = \eta_1 (\eta_1 + 2 \eta_2)/(\eta_1 + \eta_2)$. 

The cross-sectional area $A(\phi, d)$ is given by 
\begin{equation}
    A(\phi, d) = \Azero [1- \kappa (\phi, d)]
    \label{eq:def_A}
\end{equation}
with the cross-sectional area $\Azero = \alpha H W$ for $\phi=0$. 
Here, $\kappa (\phi, d)$ is the reduction rate of the cross-sectional area by the groove, 
\begin{equation}
    \kappa(\phi, d) = \left\{
    \begin{array}{lll}
        \phi \,d/(\alpha H), ~& 0 \leq d \leq \alpha H \\
        \phi, ~& \alpha H < d \leq H
    \end{array}
    \right. \ .
    \label{eq:def_kappa}
\end{equation}
Substituting Eqs.~\eqref{eq:sigma_yy} and \eqref{eq:def_A} into Eq.~\eqref{eq:eom_ldl} and taking the limit of $\mrd y \rightarrow 0$, we obtain 
\begin{equation}
    \rho (1- \kappa) \ddot{q}_y (y, t) = (1- \kappa) \sbr{ E_1 \frac{\partial^2 q_y(y,t)}{\partial y^2} + \etat \frac{\partial^2 \dot{q}_y (y,t)}{\partial y^2} } 
    - \frac{\mu(\dot{q}_y (y,t)) \Pext}{\alpha H} \ .
    \label{eq:eom_ldl_continuum}
\end{equation}
The boundary conditions are given by $\partial q_y(L,t)/\partial y = 0$ for the free boundary at $y=L$ and $q_y(0, t) = U$ for the fixed boundary at $y=0$. 

We set $t=0$ just after the bulk sliding, where the friction coefficient is given by $\mu = \muK$. 
When a precursor slip occurs with the normalized slip area $\Stilde$ for $U > 0$, the friction coefficient is given by $\mu = \muS$ in the region $0 \leq y/L \leq \Stilde$, because the slip distances of the precursors are significantly smaller than that in bulk sliding. 
In the other regions, $\mu$ remains $\muK$ due to the frictional stress drop after the bulk sliding. 
This is confirmed by direct numerical calculations of Eq.~\eqref{eq:eom_ldl_continuum} and qualitatively consistent with the results in Sect.~\ref{subsec:fem_simulation}. 
For sufficiently slow driving with $\ddot{q}_y \thickapprox 0$ and $\dot{q}_y \thickapprox 0$, the quasi-static solution of $q_y$ in Eq.~\eqref{eq:eom_ldl_continuum} is analytically derived as described in Appendix~\ref{apsec:quasi-static}. 
In this quasi-static solution $q_\mathrm{a} (y)$, $\Stilde$ is given as an increasing function of $U$. 

We conduct a stability analysis based on Eq.~\eqref{eq:eom_ldl_continuum} following the procedure in the previous studies~\cite{Otsuki2013, IwashitaSciRep2023}. 
Substituting $q_y(y,t) = q_\mathrm{a}(y) + \delta q(y,t)$ into Eq.~\eqref{eq:eom_ldl_continuum} with the perturbation $\delta q(y,t)$, we obtain the equation for $\delta q(y,t)$ as 
\begin{equation}
    \rho (1- \kappa) \delta\ddot{q} (y, t) = (1- \kappa) \sbr{ E_1 \frac{\partial^2 \delta q(y,t)}{\partial y^2} + \etat \frac{\partial^2 \delta\dot{q} (y,t)}{\partial y^2} } 
    - \frac{(\muS - \muK) \Pext}{\vc \alpha H} \delta \dot{q}(y,t) \ .
    \label{eq:eom_deltaq}
\end{equation} 
Note that $\delta q(y,t)$ has a non-zero value in the region $0 < y/L < \Stilde$, and $\delta q(y,t)$ remains zero in the other region due to static friction. 
Since the perturbation $\delta q(y,t)$ is zero for $y = 0$ and $\Stilde < y/L < 1$, $\delta q(y,t)$ is expressed as 
\begin{equation}
    \delta q(y,t) = \sum_m q_m e^{\lambda_m t} \sin{k_m \xi} \ ,
    \label{eq:deltaq_e_sin}
\end{equation}
where $m$ is a positive integer, $q_m$ is a constant, $\lambda_m$ is the eigenvalue of the time evolution operator with $k_m = m\pi$ and $\xi = y/(\Stilde L)$. 
Substituting Eq.~\eqref{eq:deltaq_e_sin} into Eq.~\eqref{eq:eom_deltaq}, multiplying by $2 \sin{k_n \xi}$ with positive integer $n$, and integrating in $0 < y < \Stilde L$, we obtain 
\begin{equation}
    (1- \kappa)  \rho L^2 \lambda_m^2 + (1- \kappa)  E_1 \frac{k_m^2}{\Stilde^2} + (1- \kappa) \etat  \frac{k_m^2}{\Stilde^2} \lambda_m  
    - \frac{(\muS - \muK) \Pext L^2}{\vc \alpha H} \lambda_m  = 0 \ .
    \label{eq:lambda_m}
\end{equation}

The perturbation $\delta q(y,t)$ is unstable in the case of $\Re \lambda_1 > 0$ and $\Im \lambda_1 \neq 0$. 
The latter condition, $\Im \lambda_1 \neq 0$, induces the oscillatory motion. 
However, the backward motion of the oscillation reduces the frictional stress, and the local slip stops when it becomes smaller than the local maximum static frictional stress, which causes the reduction of the slip area $\Stilde$.
This oscillatory instability corresponds to the BRP. 
The BRP continues in a certain region of $\Stilde$ because the frictional stress increases again by the drive, and intermittent slip events are observed until the perturbation develops and causes bulk sliding in the case of $\Re \lambda_1 > 0$ and $\Im \lambda_1 = 0$. 
In Eq.~\eqref{eq:lambda_m}, we find that the stability conditions of the system are generally determined by the competition between the viscosity represented by the third term and velocity-weakening friction represented by the fourth term on the left-hand side. 
These terms are considered as the stabilizing and destabilizing factors, respectively. 
The stabilizing factor decreases due to $\Stilde^{-2}$ in the third term as the precursor slip area $\Stilde$ increases. 
When $\Stilde$ reaches the critical area $\Sc$, the destabilizing factor overwhelms the stabilizing factor, and the perturbation $\delta q(y,t)$ becomes unstable. 
Therefore, $\Stilde$ increases rapidly just after reaching $\Sc$, as shown in Fig.~\ref{fig:stilde_U}, and bulk sliding occurs. 
The viscous term is proportional to $1-\kappa$. 
The velocity-weakening friction term is proportional to the load on the top of the block but independent of $\kappa$. 
Thus, if $\kappa (\phi, d)$ increases by increasing $\phi$ and $d$, the viscosity becomes effectively smaller, which leads to the decrease of $\Sc$. 

As $\Stilde$ increases, the mode with $m=1$ in Eq.~\eqref{eq:deltaq_e_sin} becomes unstable first, which determines $\Sc$. 
By definition, the maximum value of $\Sc$ does not exceed $1$, and for $\Sc < 1$, $\Sc$ satisfies 
\begin{equation}
    \pi^2 (1- \kappa) \etat \Sc^{-2} + 2 \pi (1- \kappa) L \sqrt{ \rho E_1 } \Sc^{-1} 
    = \frac{\paren{ \muS - \muK } \Pext L^2}{\vc \alpha H} \ ,
    \label{eq:sc_ver1}
\end{equation}
which is derived from Eq.~\eqref{eq:lambda_m}. 
Therefore, $\Sc$ is given by 
\begin{equation}
    \Sc = \min(\Sc^*,1) \ ,
    \label{eq:sc_real}
\end{equation}
where $\min(a,b)$ is a function that takes the smaller value between $a$ and $b$, and $\Sc^*$ is the solution of Eq.~\eqref{eq:sc_ver1} given by 
\begin{equation}
    \Sc^* = 
        \dfrac{\pi \etat}{L \paren{ - \sqrt{\rho E_1} + \sqrt{ \rho E_1 + \frac{\muS - \muK}{1-\kappa} \, \frac{\Pext \etat}{\vc \alpha H} } } } \ .
\end{equation}
For $\Sc \ll 1$, $\Sc$ is approximately given by 
\begin{equation}
    \Sc \simeq \pi \cbr{ \frac{[1-\kappa(\phi, d)]\alpha}{\muS - \muK} \, \frac{\etat \vc}{\Pext H} }^{\frac{1}{2}} \frac{H}{L} \ .
    \label{eq:sc_ver3}
\end{equation}
This result indicates that the normalized critical area of the precursor slip $\Sc$ is a decreasing function of $\Pext$ and the size of grooves because $\kappa (\phi, d)$ in Eq.~\eqref{eq:sc_ver3} increases with $\phi$ and $d$, as described in Eq.~\eqref{eq:def_kappa}. 
These analytical results are qualitatively consistent with those of the FEM simulations shown in Fig.~\ref{fig:sc_d}. 

The macroscopic static friction coefficient $\muM$ can be analytically derived in our simplified model~\cite{Otsuki2013, IwashitaSciRep2023}. 
Since the ratio of the local frictional stress to bottom pressure is $\muS$ in the slip region and $\muK$ in the static region, $\muM$ just before the bulk sliding is given by 
\begin{equation}
    \muM = \muK + (\muS - \muK) \Sc \ .
    \label{eq:muM_Sc}
\end{equation}
This result is qualitatively consistent with the FEM simulations shown in Fig.~\ref{fig:muM_Sc_tp_L}, where Eq.~\eqref{eq:muM_Sc} is represented by the solid line. 

Substituting Eq.~\eqref{eq:muM_Sc} into Eq.~\eqref{eq:sc_ver3}, we obtain
\begin{equation}
    \muM - \muK \simeq \pi \cbr{ (\muS - \muK) [1-\kappa(\phi, d)] \alpha\, \frac{\etat \vc}{\Pext H} }^{\frac{1}{2}} \frac{H}{L} \ .
    \label{eq:muM_Pext_kappa}
\end{equation}
This equation, together with Eq.~\eqref{eq:def_kappa}, implies that the macroscopic static friction coefficient $\muM$ is a decreasing function of $\Pext$, $\phi$ and $d$. 
The analytical results are qualitatively consistent with the FEM simulations shown in Fig.~\ref{fig:muM_d}. 

\begin{figure}[b]
    \centering
    \includegraphics[scale=0.7]{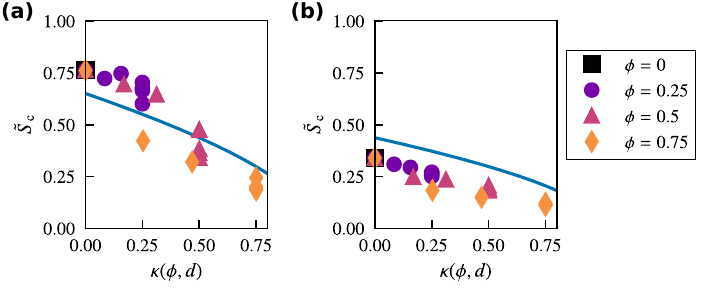}
    \caption{
        Normalized critical area of precursor slip $\Sc$ against $\kappa (\phi, d)$ for different values of $\phi$ and $d$ with (a)~$\Pext/E=0.003$ and (b)~$\Pext/E=0.006$. 
        The symbols represent the results of the FEM simulations. 
        The solid lines represent the analytical results given by Eq.~\eqref{eq:sc_real} with $\alpha=0.2$. 
    }
    \label{fig:kappa_Sc}
\end{figure}

\begin{figure}[b]
    \centering
    \includegraphics[scale=0.7]{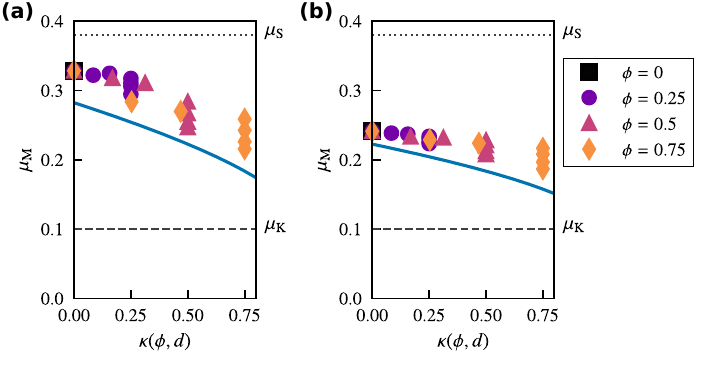}
    \caption{
        Macroscopic static friction coefficient $\muM$ against $\kappa (\phi, d)$ for different values of $\phi$ and $d$ with (a)~$\Pext/E=0.003$ and (b)~$\Pext/E=0.006$. 
        The symbols represent the results of the FEM simulations. 
        The solid lines represent the analytical results given by Eqs.~\eqref{eq:sc_real} and \eqref{eq:muM_Sc} with $\alpha=0.2$. 
        The dotted and dashed lines indicate $\muS$ and $\muK$, respectively. 
    }
    \label{fig:kappa_muM}
\end{figure}

Equations~\eqref{eq:sc_ver3} and \eqref{eq:muM_Pext_kappa} indicate that $\Sc$ and $\muM$ for different $\phi$ and $d$ are scaled by the reduction rate of cross-sectional area $\kappa (\phi, d)$. 
Figures~\ref{fig:kappa_Sc} and \ref{fig:kappa_muM} respectively show $\Sc$ and $\muM$ obtained from the FEM simulations against $\kappa (\phi, d)$. 
Both $\Sc$ and $\muM$ are scaled by $\kappa (\phi, d)$ and decrease with increasing $\kappa (\phi, d)$. 
The solid line in Fig.~\ref{fig:kappa_Sc} represents the analytical result given by Eq.~\eqref{eq:sc_real}. 
The solid line in Fig.~\ref{fig:kappa_muM} represents the result given by Eqs.~\eqref{eq:sc_real} and \eqref{eq:muM_Sc}. 
Here, we set $\alpha = 0.2$ to semi-quantitatively reproduce the results of the FEM simulations in Sect.~\ref{subsec:fem_simulation}. 
The deformation before bulk sliding is significant only in the region $z/H < 0.5$, as shown in Appendix~\ref{apsec:acc_cross-section}. 
This result is consistent with the estimate of $\alpha = 0.2$. 
The numerical results of FEM are semi-quantitatively consistent with the theoretical analysis. 

These results explain the decreases of $\muM$ and $\Sc$ with the increases of $\phi$ and $d$ in Figs.~\ref{fig:muM_d} and \ref{fig:sc_d}. 
Here, $\phi$ represents the decrease in the contact area of the pillars, and $d$ represents the decrease in the size of the main body. 
These values determine the reduction rate of the cross-sectional area $\kappa (\phi, d)$. 
Thus, it can be concluded that $\muM$ and $\Sc$ decrease with increasing $\phi$ or $d$ because of the decrease in effective viscoelasticity due to the reduction of the cross-sectional area, leading to the decline of the stability and bulk sliding with a smaller size of the precursor slip.

\section{Discussion} \label{sec:discuss}

Generally, grooves on friction surfaces are designed to control lubrication properties in wet conditions. 
It is considered that the friction coefficient at the wet interface increases with the width and depth of the grooves, because they can eject more lubricant from the friction interface~\cite{Li2004, Li2005, Li2006, Yamaguchi2012}. 
However, this study reveals that the groove size also affects friction in dry conditions. 
The static friction coefficient in dry conditions decreases with increases in groove width and depth. 
This is opposite to the usual consideration for the friction in the wet case. 
Even in wet conditions, the friction force at the solid-solid interface determines the total friction force after the ejection of the lubricant. 
These results should aid in improving the design of sliding interfaces with grooves for both wet and dry conditions. 

The influence of the groove shape on friction in dry conditions has recently been investigated based on different models, where the effect of grooves is represented by the spatial distribution of the local friction coefficient~\cite{Capozza2015, Costagliola2016, Costagliola2017, Maegawa2017, Costagliola2018, Costagliola2022IJSS, Berardo2019, Balestra2022}.
These previous works report a decrease in the static friction coefficient by forming longitudinal grooves, which is consistent with our results. 
However, the effect of the depth of the longitudinal grooves is ignored in their models, while our study is based on a realistic 3D system and reveals its importance. 
Moreover, previous studies have investigated different patterns of grooves including transversal grooves. 
Their results have suggested that complex shapes in the friction surface used in various industrial products such as shoe soles and tires and adopted on the surface of living things such as snakes~\cite{Baum2014BJN, Baum2014TL, Costagliola2017} affect the frictional properties. 
Therefore, the extension of our study to these complex shapes will lead to more efficient guiding principles for groove design. 

In this study, the parameter values for virtual materials are adopted to reduce the computational load, which does not correspond to those for real materials. 
These are selected to compare the results with those in previous simulations~\cite{Otsuki2013, IwashitaSciRep2023}. 
However, the mechanism of changes in the friction coefficient revealed by our theory in Sect.~\ref{subsec:theory_analysis} is universal and independent of specific parameter values. 
The numerical results for flat friction surfaces in Ref.~\citenum{Otsuki2013} have been reproduced in experiments on PMMA~\cite{Katano2014}. 
The occurrence of the quasi-static precursor (slow slip event, SSE) and the dependence of its destabilization on pressure are also confirmed in an experiment on PMMA~\cite{Selvadurai2017}. 
Therefore, we expect our results will be experimentally verified in future work. 

\section{Conclusion} \label{sec:conclusion}

Friction surfaces of products such as shoe soles, tires, and sliding parts of machines have grooves. 
Several studies on grooves have focused on controlling lubrication properties via their design. 
However, it has been empirically known that grooves affect friction even in dry conditions, although no theoretical explanation exists. 
In this study, we have performed numerical simulations of viscoelastic objects using 3D FEM to clarify the effect of longitudinal grooves on static friction in a dry condition. 
We have revealed that the static friction coefficient is a decreasing function of the groove size, and that precursor slip occurs before bulk sliding. 
The static friction coefficient is scaled by the normalized critical area of the precursor slip. 
Based on the simplified model, we have theoretically derived the equation for the static friction coefficient depending on the groove size. 
The theoretical result indicates that the static friction coefficient decreases with the reduction rate of the cross-sectional area in the viscoelastic object. 
The decrease in the cross-sectional area reduces the effective viscosity, which enhances the instability of the precursor slip and decreases the static friction coefficient. 
Our results provide new guiding principles for groove design for static friction control beyond the empirical laws for both wet and dry conditions. 
Investigation of different types of grooves and their effect on dynamic friction will be the subject of future studies.

\backmatter

\section*{Data Availability}
The data that support the findings of this study are available from the corresponding author upon reasonable request. 

\bibliography{sn-bibliography}

\bmhead*{Acknowledgments}
This research used computational resources of Fugaku at the RIKEN Center for Computational Science (Project: hp220372), Yukawa-21 at YITP, Kyoto University, SQUID at the Cybermedia Center, Osaka University, ohtaka and kugui at ISSP, the University of Tokyo, the supercomputers at RCCS, Okazaki, Japan (Project: 23-IMS-C126), and JSS3 at JAXA. 
We would like to thank Editage (www.editage.com) for English language editing. 

\bmhead*{Funding}
This work was supported by the Japan Society for the Promotion of Science KAKENHI (Grant Numbers JP21H01006, JP22KJ2190, JP23K03248, and JP23K03252). 

\section*{Author information}
\bmhead*{Contributions}
All authors contributed to the study conception and design. 
The FEM simulation and analysis based on a simplified model are performed by WI. 
The first draft of the manuscript was written by WI, and all authors commented on previous versions of the manuscript. 
All authors read and approved the final manuscript. 

\section*{Ethics declarations}
\bmhead*{Competing interests}
The authors declare no competing interests. 

\section*{Supplementary information}
The online version contains supplementary material.

\newcounter{figmemo}\setcounter{figmemo}{\value{figure}} 
\begin{appendices}
\renewcommand{\thefigure}{\arabic{figure}} 
\setcounter{figure}{\value{figmemo}} 

\section{Results for \texorpdfstring{$L/H=2$}{L/H=2}} \label{apsec:dependence_L}

\begin{figure}[b]
    \centering
    \includegraphics[scale=0.7]{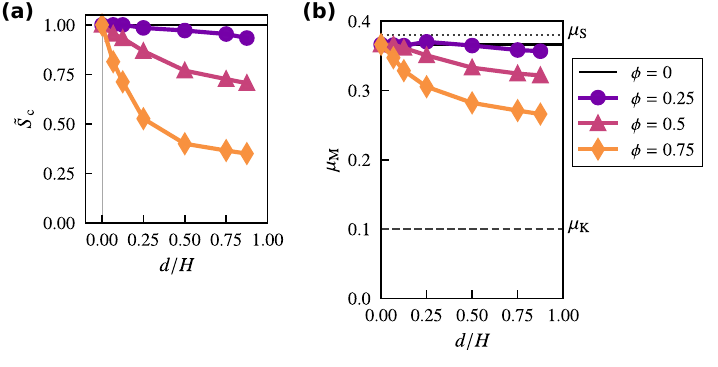}
    \caption{
        (a)~Normalized critical area of precursor slip $\Sc$ and (b)~macroscopic static friction coefficient $\muM$ against $d$ for different values of $\phi$ with $L/H=2$ and $\Pext/E=0.003$. 
        The dotted and dashed lines indicate $\muS$ and $\muK$, respectively. 
    }
    \label{fig:muM_Sc_d_L2}
\end{figure}

\begin{figure}[t]
    \centering
    \includegraphics[scale=0.7]{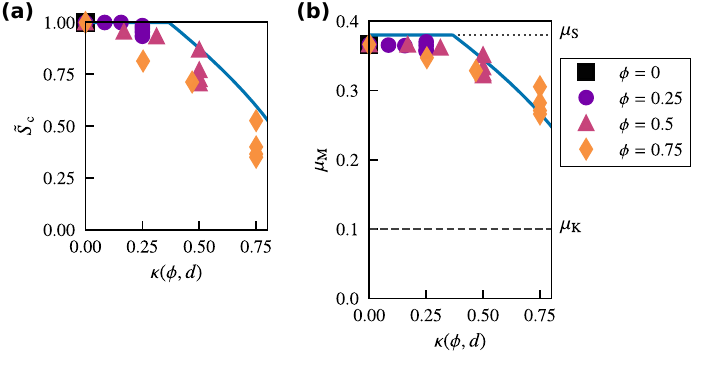}
    \caption{
        (a)~Normalized critical area of precursor slip $\Sc$ and (b)~macroscopic static friction coefficient $\muM$ against $\kappa (\phi, d)$ for different values of $\phi$ and $d$ with $L/H=2$ and $\Pext/E=0.003$. 
        The solid lines represent the analytical results of Eqs.~\eqref{eq:sc_real} and \eqref{eq:muM_Sc} with $\alpha=0.2$. 
        The dotted and dashed lines indicate $\muS$ and $\muK$, respectively. 
    }
    \label{fig:muM_Sc_kappa_L2}
\end{figure}

In this appendix, we show the results for $L/H=2$. 
Figure~\ref{fig:muM_Sc_d_L2}a and b respectively show $\Sc$ and $\muM$ against $d$ for different values of $\phi$ with $\Pext/E=0.003$. 
We find that $\Sc$ and $\muM$ are decreasing functions of $\phi$ and $d$, which is consistent with the results for $L/H=4$, as shown in Figs.~\ref{fig:muM_d} and \ref{fig:sc_d}. 
Compared to Figs.~\ref{fig:muM_d}a and \ref{fig:sc_d}a for the identical pressure, we find that $\Sc$ and $\muM$ increase with decreasing $L$. 

Figure~\ref{fig:muM_Sc_kappa_L2}a and b respectively show $\Sc$ and $\muM$ against $\kappa (\phi, d)$ with $\Pext/E=0.003$. 
We find that $\Sc$ and $\muM$ are decreasing functions of $\kappa (\phi, d)$. 
The solid lines show the analytical results for $\alpha=0.2$ based on Eqs.~\eqref{eq:sc_real} and \eqref{eq:muM_Sc}. 
It is shown that $\kappa (\phi, d)$ for $\alpha=0.2$ can scale $\Sc$ and $\muM$ even for the system with $L/H=2$. 
The decreases in $\Sc$ and $\muM$ by increasing $L$ is consistent with Eqs.~\eqref{eq:sc_ver3} and \eqref{eq:muM_Pext_kappa} for the theoretical analysis in Sect.~\ref{subsec:theory_analysis}.

\section{Dependence on Number of Pillars \texorpdfstring{$n_x$}{nx} in Friction Surface} \label{apsec:dependence_nx}

\begin{figure}[b]
    \centering
    \includegraphics[scale=0.7]{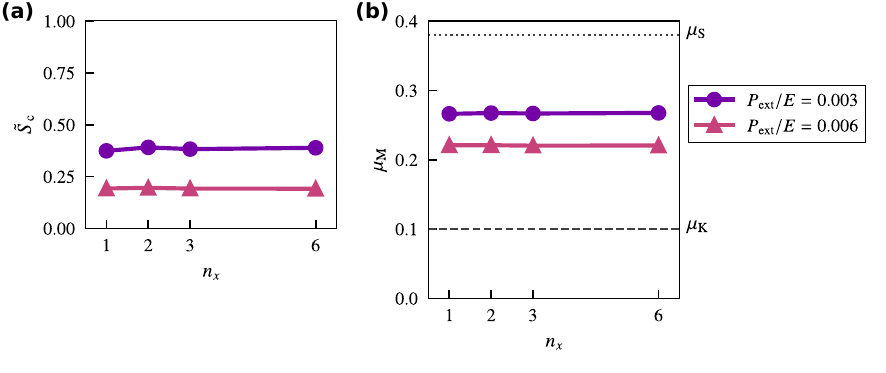}
    \caption{
        (a)~Normalized critical area of precursor slip $\Sc$ and (b)~macroscopic static friction coefficient $\muM$ against $n_x$ for $\Pext/E=0.003$ and $0.006$ with $L/H=4$, $d/H=0.5$, and $\phi=0.5$. 
        The dotted and dashed lines represent $\muS$ and $\muK$, respectively. 
    }
    \label{fig:muM_Sc_nx}
\end{figure}

In this appendix, we verify the dependence of the $\Sc$ and $\muM$ on the number of pillars of the block $n_x$ for $L/H=4$, $d/H=0.5$, and $\phi=0.5$.
Figure~\ref{fig:muM_Sc_nx}a and b show $\Sc$ and $\muM$ against $n_x$ for $\Pext/E=0.003$ and $0.006$, respectively. 
In Fig.~\ref{fig:muM_Sc_nx}, we find that $\Sc$ and $\muM$ are almost independent of $n_x$. 
According to the theoretical results given by Eqs.~\eqref{eq:sc_ver3} and \eqref{eq:muM_Pext_kappa}, $\Sc$ and $\muM$ depend on the reduction rate of the cross-sectional area $\kappa (\phi, d)$, which is independent of $n_x$, as described in Eq.~\eqref{eq:def_kappa}.
This analytical result explains the behaviors of $\Sc$ and $\muM$ in Fig.~\ref{fig:muM_Sc_nx}.

\section{Acceleration Distribution} \label{apsec:acc_cross-section}

\begin{figure}[H]
    \centering
    \includegraphics[scale=0.7]{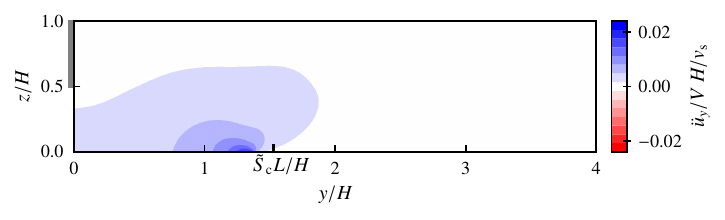}
    \caption{
        Spatial distribution of acceleration along the $y$ direction, $\ddot{u}_y$, on cross-section at $x=0.5 W$ for $\Pext/E=0.003$, $d/H=0.5$, and $\phi=0.5$ at $U/L=5.43 \times 10^{-3}$, just before bulk sliding.
        The gray thick line for $y/H=0$ and $0.5 \leq z/H \leq 1$ indicates the position of the rigid plate. 
    }
    \label{fig:cmap_cs}
\end{figure}

Figure~\ref{fig:cmap_cs} demonstrates the spatial distribution of the $y$-direction acceleration $\ddot{u}_y$ for $\Pext/E=0.003$, $d/H=0.5$, and $\phi=0.5$ in the cross-section at $x=0.5 W$ with $U/L=5.43 \times 10^{-3}$, just before bulk sliding. 
The critical length in the $y$ direction of the precursor slip is estimated to be $\Sc L$, because the precursor slip propagates uniformly in the $x$-direction, as shown in Fig.~\ref{fig:cmap_precursor}.
The acceleration around the region of $y/H=0$ and $0.5 \leq z/H \leq 1$ is small, because the displacement is fixed at the rigid plate position at the region. 
Moreover, the acceleration is negligible in the static region at the bottom ($z = 0$) with $\sigmaf/p \thickapprox \muK$ for $\Sc L \leq y \leq L$, as shown in Fig.~\ref{fig:cmap_tp}a. 
Therefore, the region with significant acceleration before bulk sliding is concentrated in the region for $0 \leq y \leq \Sc L$ near the bottom, as shown in Fig.~\ref{fig:cmap_cs}. 

\section{Quasi-static Solution} \label{apsec:quasi-static}

Using the equation of motion~\eqref{eq:eom_ldl_continuum}, boundary conditions, and assumption of friction coefficient $\mu$ in Sect.~\ref{subsec:theory_analysis}, we obtain the quasi-static solution $q_0(y)$ at $U=0$ as 
\begin{equation}
    q_0(y) = \frac{\muK \Pext}{2(1-\kappa)E_1 \alpha H} \paren{y^2 - 2Ly} \ .
    \label{eq:qzero}
\end{equation}
The quasi-static solution $q_\mathrm{a}(y)$ for $U>0$ is obtained as 
\begin{equation}
    q_\mathrm{a}(y) = \left\{
    \begin{array}{cc}
        q_1(y), & ~~~0 \leq y \leq \Stilde L \\
        q_0(y), & ~~~\Stilde L \leq y \leq L
    \end{array}
    \right. \ .
    \label{eq:q_a}
\end{equation}
With the assumption of $\mu$ and connectivity condition at $y = \Stilde L$, $q_1(\Stilde L) = q_0(\Stilde L)$ and $\mrd q_1 (\Stilde L)/\mrd y = \mrd q_0 (\Stilde L)/\mrd y$, we obtain $q_1$ as 
\begin{equation}
    q_1(y) = q_0(y) + \frac{(\muS - \muK) \Pext}{2(1-\kappa) E_1 \alpha H} \paren{y^2 - 2 \Stilde L y + \Stilde^2 L^2} \ .
    \label{eq:qone}
\end{equation} 
To satisfy the boundary condition at $y/L=0$, $\Stilde$ is derived as 
\begin{equation}
    \Stilde = \sbr{ \frac{2(1-\kappa) E_1 \alpha H U}{(\muS-\muK) \Pext L^2} }^\frac{1}{2} \ .
    \label{eq:Stilde}
\end{equation}

\end{appendices}

\end{document}